\documentclass[aps,prd,nofootinbib,twocolumn]{revtex4-1}
\usepackage[english]{babel}
\usepackage[autostyle, english = british]{csquotes}
\usepackage[CJKbookmarks, pdftex, bookmarksnumbered, bookmarksopen, colorlinks, citecolor=blue, linkcolor=blue]{hyperref}
\usepackage{amsmath,amssymb}
\usepackage{graphicx}
\usepackage{enumitem}
\setlist[enumerate]{topsep=0pt,parsep=-1mm,leftmargin=5mm,}
\def\be{\begin{equation}}
\def\ee{\end{equation}}

\usepackage{color}  
\RequirePackage[dvipsnames,usenames]{xcolor}

\newcommand{\PD}{{\cal{D}}}

\begin{document}

\title{\large Why you do not need to worry about the standard argument\\ that you are a Boltzmann brain}

\author{Carlo Rovelli${}^{abcd}$, David Wolpert${}^{d}$}
\affiliation{Aix-Marseille University, Universit\'e de Toulon, CPT-CNRS, F-13288 Marseille, France.}
\affiliation{Department of Philosophy and the Rotman Institute of Philosophy, 1151 Richmond St.~N London  N6A5B7, Canada}
\affiliation{Perimeter Institute, 31 Caroline Street N, Waterloo ON, N2L2Y5, Canada} 
\affiliation{Santa Fe Institute, 1399 Hyde Park Road Santa Fe, New Mexico 87501, USA}

\begin{abstract} 
\noindent 
Are you, with your perceptions, memories and observational data, a Boltzmann brain, namely a fleeting statistical fluctuation out of the thermal equilibrium of the universe?  Arguments are given in the literature  claiming that this bizarre hypothesis needs to be considered seriously, that all of our data about the past is actually a mirage. We point to a difficulty in these arguments.  They are based on the dynamical laws and on statistical arguments, but they disregard the fact that we infer the the dynamical laws \textit{presupposing} the reliability of our data records about the past.  Hence the reasoning in favor of the Boltzmann brain hypothesis contradicts itself, relying on the reliability of our data about the past to conclude that that data is wrong. More broadly, it is based on incomplete evidence.  Incomplete evidence notoriously leads to false conclusions. 

\end{abstract}

\maketitle

\section{Boltzmann brains}

\noindent Boltzmann brain, or b-brain for short, is the name given to a phenomenon that is {\em in principle} possible in statistical mechanics. Imagine a large statistical system formed by a mixture of particles of different kinds that remains in  thermal equilibrium for an arbitrarily long span of time.   According to statistical mechanics, there are fluctuations at thermal equilibrium, and in principle all configurations can be reached by such fluctuations with enough time available. Consider one of these random fluctuations giving rise --just by chance-- precisely to a brain like yours, containing all the memories, the information and the perceptions that you have right now, fluctuating for a brief moment into existence. That would be a b-brain (see for instance \cite{Albrecht2004,Linde2007,Aguirre2012}).  That brain will have exactly your worldview, your memories, your perceptions.   It would know and feel precisely what you know and the way you feel right now.  Now, how do you know that you are actually not such a b-brain?   

A simple answer is that the probability for such a fluctuation to occur is fantastically small, and the expected time for this to happen is fantastically long, colossally longer than the current age of the universe. But recent literature has questioned this simple answer, presenting an argument meant to show that current science makes the doubt of being a b-brain harder to dispel. Some have maintained that the laws of physics must be able to suppress this possibility.  Here, we point out a weak point of this argument from the recent literature (for a similar point, see \cite{Myrvold216}). 

To start, let's see what this arguments is.   Suppose we start from these two ingredients and nothing else:
\begin{enumerate}[nosep, label=(\roman*)]
\item The set $\cal L$ of all the mechanical Laws of physics that we know.  These are symmetric under time inversion. 
\item The set $\cal D$ of all present (say at time $t=0$) observations (or Data) that we  have (this includes all records and recorded data existing in the present, including memories).
\end{enumerate} 
What can then we say about the world, based on our data $\cal D$?  The set of laws $\cal L$ admits a large family of solutions.  Let us 
restrict ourselves to these solutions to those compatible with $\cal D$, in the sense that $\cal D$ has non-infinitesimal  likelihood
under those solutions. Ignoring the other solutions, for simplicity, let us assign equal probability to all these solutions that we restrict ourselves to. 

Now, the dynamical laws (with the equiprobability assumption) imply the Boltzmann's H-theorem, which says that entropy increases both into the past and into the future from any given time at which entropy is known to have some specific low value. This theorem is time-symmetric.  The standard formulation of the second law of thermodynamics  says that we are in the future of such a special point. (However, see~\cite{Scharnhorst2024}.)  Note though that the most probable situation, given the current observations, is that we happen to be \textit{at} that special point, not in its future. In other words, the most probable situation is that we are just an entropy fluctuation. And this is to say that we are a b-brain. 

One may respond to this argument that we also know the second law of thermodynamics, and from this we know that entropy grows with $t$, i.e., that the special point with low given entropy was in our past. But our knowledge of the second law arose from consideration of our data records about the past: how do we know that these are data records of the past and aren't {\em themselves} due to a fluctuation?   

We commonly assume that entropy was in fact low in the past and has been growing since.   But assuming that the entropy in the past was (sufficiently) low amounts to a very  improbable assumption, with respect to the equiprobability of the solutions compatible with $\cal D$.   On the basis of this equiprobability, the data $\cal D$ are more likely to be the result of a fluctuation than to be the result of low entropy in the distant past.   This is simple to show: the entropy today is higher than this assumed past entropy and entropy is precisely likelihood in this sense.    Low past entropy, in this logic, is not an account of why the present is out of equilibrium: it is additional in-probability added to how special is the present. 

Additionally, if the set of laws $\cal L$ admit thermalization in the distant future and the universe settles into a long living equilibrium state (Boltzmann's ``thermal death"), then even if the universe started off in a low entropy state, still $\cal D$ can be realized not once but an infinite number of times in the distant future, making it more reasonable to infer that an observation of $\cal D$ is one of these fluctuations, rather than something realized in the initial thermalization transient.    
These arguments appear to converge to showing that we should take the hypothesis of being b-brain seriously. 

Yet, there is something strident and unconvincing in all this.  Is this just a resistance due to our wrong naive intuition, or is there a real problem in the argument above?   We think that the second case is the correct one, and here we point out where the difficulty is.   

In this paper we show that the standard argument in favor the b-brain hypothesis contradicts itself.
Our argument is semi-formal only, and we are not overly scrupulous of the precise Bayesian foundations of the philosophy
of science. Our goal is instead to highlight a (fatal, we think) flaw in the standard argument, while clarifying why
that flaw doesn't apply to our current understanding of the laws of the universe.
 
Before going to that though, there is a preliminary ambiguity to dispel.   It is different to argue that something is unlikely, even extremely unlikely, from arguing that it is impossible.  Nothing is impossible unless it violates logic. The wildest scenarios are possible, including that on the dark side of the moon little red dragons live and hide very well, that the entire universe is a dream of my single existing soul, or that elves are among us and well hidden.   This is the radical skeptical thesis, commonly restated in philosophy, which is correct.  Nobody put it more nicely than the Zhuang Zi \cite{Ziporyn2020} in the famous apologue of the dream of the butterfly: Zhuang Zi dreamt of being a butterfly happily flying around among flowers. When he woke up, he puzzled: ``How can I know if I am Zhuang Zi who dreamt of being a butterfly, or I am a butterfly dreaming of being Zhuang Zi?".   Everything is possible.   

But all these possibilities, precisely because they are innumerable and of any possible sort, are typically dismissed in science (they are lovely for poetry): the standard view is that they form a grey ocean of nonsense, all equally irrelevant, unless something makes one of them plausible.   There is no totally rigorous argument to exclude {\em any} of them. In fact, no data that we could acquire could confirm that Zhuang Zi is a butterfly dreaming --- nor could any data refute that hypothesis, establishing that he is not just such a butterfly. This is why such a hypothesis is consigned to the ``grey ocean'', under the standard view of science. In this same, trivial sense, being a b-brain cannot be rigorously excluded, but rather would be consigned to the ``grey ocean'' under the standard view of science. The puzzle of b-brains is not about the abstract \emph{possibility} of being a fluctuation: it is about the \emph{probability} of being a fluctuation.  It is not about what is logically possible: it is about what we say about how probable it is.  What the above argument is supposed to argue for is that it is likely that you are a Boltzmann brain.  

More generally, our knowledge of reality is never certain (we could always be the dream of a butterfly).  Our knowledge of the universe is based on a number of guesses that stay together coherently and that we see sufficiently confirmed to tentatively take them as reliable.  Our knowledge is never certain, but it can be very reliable:  I am very confident that if I let a stone free, it will fall down, not up.

\section{The flaw in the standard argument for a b-brain}

So, let us return to the argument.  The logic of the argument above is correct: given the premises (i) (the laws $\cal L$) and (ii) (The present data of observation  $\cal D$), it does follow that I am likely a b-brain ($\cal B$).  Let us formalize this result by writing that the probability $P({\cal B}|{\cal L},{\cal D})$ of you being a b-brain ($\cal B$), given the laws of physics $\cal L$ and the data $\cal D$ observed in the present, is near unit.   We write this as 
\be
({\cal D},{\cal L})\to {\cal B}.
\ee
The problem we intend to point out is not in the logic of the argument: it is in the premises $\cal L$ and $\cal D$.  And it is not either than (i) and (ii) are individually wrong.   It is that we \textit{infer} $\cal L$ \textit{from} $\cal D$ --- but that inference in turn invokes those very
laws $\cal L$ we wish to infer. (Specifically, it relies on our using the second law of thermodynamics~\cite{Wolpert2023}.) So we have circular
reasoning. 

Viewed differently, Both (i) and (ii) are right.    And yet, the argument is incorrect.  The problem can be seen as
arising because (i) and (ii) are \emph{incomplete}, and likelihood derived from incomplete premises can be strongly misleading.  

To show that this is the case, let us first make a general observation.   From a certain amount of information $I$ that we judge reliable, we can deduce some consequences.  If we restrict to a subset $i\subset I$ of the  information, we may deduce some consequences that are clearly false in the light of $I$.  For instance, suppose that the information $i$ is a video of a closed camera circuit, from which we see that John was killed at 6pm in his house, and at 6pm Bob, upset because John had kissed his girlfriend, was in the same house, and forensic evidence shows that the gun, with Bob fingerprints, is the weapon that killed John.  This seems to be strong evidence against Bob.  But imagine that there is also a second video, where we see that at 6pm a robber had entered John's house, taken the gun from Bob's coat hanging in the entrance, and shot John while he was discussing in a civilized manner the jealously issue with his dear friend Bob.  Then the evidence on the basis of $i$ appears to be misleading, in the light of the larger evidence $I$.  A shadowy secret service could have Bob convicted by carefully hiding the second video recorded by the closed circuit camera. 

Let us put this in a bit more formal terms. The likelihood that Bob is culpable ($B$), given the evidence provided by the two tapes $T_1$ and $T_2$ together is negative:
\be
(T_1,T_2)\to ({\rm not}\ B)
\ee
And yet, the likelihood that that Bob is culpable given the evidence provided by the incomplete evidence $T_1$ alone is high 
\be
T_1\to B
\ee
(and misleading!). There is no contradiction. Incomplete evidence can suggest high probability to false conclusions! 

We are now going to argue that the two conditions (i) and (ii) are incomplete evidence, and the b-brain thesis is like Bob's culpability: a false conclusion from selective incomplete evidence.     

To this aim, let's focus on the assumption (i): the laws $\cal L$.   How do we know the laws of the universe that we claim to know?  By knowing
only some present data $\cal D$, i.e., properties of the current universe, we cannot infer these laws. To make such an inference,
we must know some data concerning the properties of the universe in the past, $\cal P$. This is because the 
laws $\cal L$ are relations between the values of the physical variables \emph{at different times},  hence we cannot know or infer anything about them unless we also know data about the world at a time different that the present.  That is, we have 
\be
({\cal P},{\cal D})\to {\cal L}.  
\ee

How do we access the past data $\cal P$?  As described in~\cite{Wolpert2023},
to do this we have to trust the reliability of \emph{records} or \emph{memories} we have at the present that concern these past data. 
Let us call $\cal R$ the assumption that such memories of past data are statistically reliable.\footnote{In the terminology of~\cite{Wolpert2023},
the results of almost all science experiments are recorded in a ``type-3'' memory. The second law can be
applied to deduce information about the past state of the universe from the current state of such a memory. Accordingly, we can formalize 
$\cal R$ as the statement that we have a type-3 memory, whose current state is specified in $\cal D$,
and that provides the information $\cal P$ about the past state of the universe. } For notational simplicity, we indicate these
memories as parts of $\cal D$, so that the combination $(\cal{D}, \cal{R})$ establishes $\cal P$.
So whenever we write $(\cal{D}, \cal{R})$, that is equivalent to writing $(\cal{D}, \cal{R}, \cal{P})$.

Now, let us consider the two cases separately: whether we make the assumption $\cal R$ or we do not.
If we do, we have 
\be
({\PD} ,{\cal R})\to {\cal L}. 
\label{questa}
\ee
But this yields the opposite of what is given by the incomplete evidence! In fact, recalling that 
the simplified notation in \eqref{questa} means that $P({\cal L}|\PD,{\cal R})$ is close to $1$, we have 
$P(~{\cal L} | \PD,{\cal R}) \simeq 0$. This in turn implies
\be
P({\cal B}|\PD,{\cal R})=P({\cal B}|\PD,{\cal R},{\cal L})P({\cal L}|\PD,{\cal R}).
\ee
The second term on the right side is large, but the first term, $P({\cal B}|\PD,\cal{R},{\cal L})$,  is near zero, because $\cal B$ contradicts $\cal R$.

Therefore, returning to our simplified notation, 
\be
(\PD,{\cal R})\to ({\cal L},\PD,{\cal R})\to ({\rm not}\ {\cal B}).
\ee
In order to have (reliable) memories or traces we need dissipation towards the future, namely increasing entropy in the time before the present \cite{Wolpert2023,Rovelli2020}, while the b-brain hypothesis assumes that entropy decreases in this time (to lead to the large fluctuation). That is, assuming that traces are reliable yields to the expect no no b-brans.  This is the precisely the case of incomplete evidence. The conclusion of b-brain is false: it is an illusion due to selected incomplete evidence. Taking account all available evidence voids it.

What about if we instead do \emph{not} assume $\cal R$?   In this case, we have no argument in support of $\cal L$. Of course $\cal B$ does not follow from $\cal D$ alone, without $\cal L$. Without an assumption about reliability of memories, we have no reasons to believe in Boltzmann brains.  Data at a single moment of time are compatible with any dynamical law. At best they characterize the state of the universe at a single time.  They are not informative about the laws of the universe.  

One may object: we can consider $\cal L$ not as a component of our knowledge, but rather as a fact of the world. That is, suppose we say: the world is truly governed by the laws $\cal L$, irrespectively from our knowledge.   What can then we deduce from this fact conjoint with the fact that we know $\cal D$?  This appears to circumvent the observation that our knowledge of $\cal L$ depends on our knowledge of records and on the assumption $\cal R$.   But this objection does not hold.  Assigning likelihood to an event is a subjective matter, not an objective matter that pertains solely to the event. By themselves, events either happen or don't.  Facts are either true of false.  To reason about likelihood can only be based on incomplete knowledge and assumptions. The only reasonable question here is the likelihood that {\em we} assign to the possibility of being of b-brains, and this only depends on what {\em we} consider reliable. Our actual considering $\cal L$ reliable is grounded into our considering records reliable, hence not being in a fluctuation.   That is, necessitates the assumption $\cal R$.

This discussion may leave a sense of confusion.  How  do we  actually know the laws?  Don't we derive them from current observations? 

The answer is yes, but not directly.   In the standard practice of science, the answer is evident: we do rely on data {\em at different times}. We do assume memories and traces to be at least to some extent reliable.  When we are at a given moment of time, how can we access data at different times? For this  we need them to be recorded and their records to be all present to us in a reliable manner.  For this, we have to assume that the records and our memory are reliable. 

The key subtle point here is that knowledge is a process, not a static deduction happening out of time, and is never directly extracted from data.  Rather, it is based on prior knowledge, constantly updated by new data.   This dynamical aspect of knowledge is what De Finetti clarified, in addition to the static probabilism of Ramsey.  We do not build knowledge from scratch. We start by assigning some prior probability $P(A)$ to an hypothesis $A$ and then we reinforce or weaken it on the basis of new data, implicitly using Bayes theorem: the updated likelywood of $A$ given new evidence $E$ is given by 
\be
P(A|E)=\frac{P(E|A)}{P(E)}P(A).
\ee
This allows us to test our hypothesis and reinforce them if they give good predictions.   The entire logic is dynamical, not statical.  It \emph{presupposes} assumptions and hypotheses. This is the way science and knowledge work in general.  

In conclusion, as far as the argument for b-branes given above is concerned, these considerations move the b-brain hypothesis back into the family of the elves that hides well: a nice story, in the vast realm of the everything and the opposite of everything being possible.   It is possible that you, my reader, are a b-brain, and this text you are reading does not exist and is instantaneously hallucinated by your neurones randomly happening to fluctuate into this perception.   But this hypothesis is as likely as you being the dream of a butterfly.  

Credibility comes progressively from coherence.  The world picture in which the currently known laws $\cal L$, the past had lower entropy, hence records are reliable, hence there was the history that we know, including the evolution that has given us a brain capable to remember, adapt, learn, and construct this very world picture, is incomplete but overall coherent.  We, as thinking beings, are ourselves dissipative systems that can perceive, think, reflect, deduce, only by being dissipative processes.  We ourselves are comprehensible as phenomena in a gradient of entropy that allows these processes. Circularity is unavoidable (we trust the second law because we rely on records that are reliable thanks to dissipation, namely to the second law itself), but is not a problem, as long as the resulting picture remains coherent, effective and predictive.  The coherence of the picture is not enhanced, but shuttered, in the b-brain hypothesis. All knowledge is ultimately hypothetical.  Some hypotheses are more coherent, explanatory and useful than others, and survive testing. The hypothesis that our records are more or less reliable is a guess that leads to a coherent and predictive worldview. It is a guess more credible than the hypothesis of being a Boltzmann brain, which depends on forcefully discarding a major component of our evidence --precisely the records of the past-- and selectively keeping a piece of knowledge derived by the discarded evidence.  All evidence can be doubted, but doubting selectively leads to shaky conclusions.

\section{Overview of the H theorem, b-brains, and the second law}

Let's refer to a hypothesis as ``data-independent'' if there is no data that can either confirm or refute it. Any
such hypothesis is outside the domain of science, or viewed equivalently, all reasoning concerning such
a hypothesis reduces to reasoning about prior probabilities (since by assumption, the likelihood function is irrelevant). 
As an example, the parable of the butterfly dreaming that it is
Zhuang Zhi is data-independent (as are most versions of solipsism, arguably). Similarly, if we
are given \textit{only} data at the present, $\cal D$, and make no further assumptions, then no such data can either confirm or
refute the b-brain hypothesis. For that matter, without any further assumptions, no data can either confirm or refute \textit{any}
of the laws of physics --- including the second law of thermodynamics. 

So all of science ultimately relies on priors, to even get to the point where there's a way for current data to have an effect on 
the conclusions reached by science.
We do not allow ourselves arbitrary freedom in setting the priors though. They must be logically consistent with one another, at a
minimum. 

In particular, suppose we assume that the laws of the universe are time-translation invariant with high probability.
Suppose we also assume that  ${\cal{D}}_1$, that part of our current data that concerns experimental tests
in the past of the microscopic, classical dynamics laws of the universe, is reliable.
Combining these two (minimal) assumptions allows us to use that current data of ours
to establish that Newton's laws, the laws of electromagnetism, etc., all hold now and also held in our past.
We can then use this conclusion to establish Boltzmann's H-theorem --- this is precisely what Boltzmann and his predecessors did, in fact.

That H-theorem tells us that entropy must increase \textit{in both directions
of time} as we move away from any special moment at which we have a given, low value of entropy~\cite{Scharnhorst2024}. So in particular, 
if we take the present to be such a special moment, with a ``given, low value of entropy''.
it implies that entropy increases into our past. This is the standard argument for the b-brain hypothesis. 

But if we accept those microscopic, classical dynamics laws of the universe, what do we think the basis could
be for our data  ${{\cal{D}}}_1$ concerning the past
to be reliable in the first place? As mentioned above, all known processes that could provide reliable memories
about the past rely on the second law~\cite{Wolpert2023,Rovelli2020}. But that law is precisely what is ruled out by
the b-brain hypothesis! So the standard argument for the b-brain hypothesis contradicts the very assumptions it relies on.

%

Of course, the foregoing does not mean that the b-brain hypothesis is false --- it just means that the standard reasoning behind that hypothesis
contradicts itself. Indeed, variants of the b-brain hypothesis don't contradict themselves the same way that the
standard version does. In particular, consider the hypothesis that there was a b-brain fluctuation in the
state of the universe exactly $500$ years ago. In other words, change the ``special moment at which we have a given, low value of entropy'' that we
use in the H-theorem from the present to a time $500$ years ago.
Under this variant of the b-brain hypothesis, entropy would have been increasing over
the past $500$ years. That would in turn mean that our memories are reliable, and so everything is consistent.

This variant of the b-brain hypothesis doesn't contradict itself, like the usual b-brain hypothesis does. However, this variant
of the b-brain hypothesis is data-independent --- no data in $\cal D$ could either confirm or refute it. 
And as we pointed out above, there are an infinite
number of hypotheses that are data-independent, which jointly constitute a ``grey mass'' of hypotheses that are
outside the purview of science. 

Do these considerations give us a reason to doubt the second law itself though? The answer is no. To see
this, simply note that  we have other data concerning experiments in the past besides ${\cal{D}}_1$. In particular,
we have data ${\cal{D}}_2$ telling us that in fact entropy \textit{decreases} into the past, not that it increases. So
if we assume that this data ${\cal{D}}_2$ is reliable, we conclude that entropy increases in time, i.e., that the
second law holds. So we get no contradiction. 

But how do we reconcile an assumption that ${\cal{D}}_1$ is reliable (so that the H-theorem holds) with
the assumption that ${\cal{D}}_2$ is reliable (so that the second law holds)? The answer is subtle. To illustrate
it, suppose that we simply replace the phrase ``$500$ years into the past'' in the variant of the b-brain hypothesis
introduced above  with ``$13$ billion years into the past''. That would transform the $500$ years-ago variant of the b-brain hypothesis into what is
commonly called the ``Past hypothesis''. \textit{That} hypothesis --- that variant of the b-brain hypothesis ---
is widely believed to be true. In fact, it is the most widely accepted explanation for {how} the second law can hold, despite
the H-theorem: one is simply changing the ``special moment at which we have a given, low value of entropy'' that we
use in the H-theorem from the present to the time of the big bang, $13$ billion years ago.

One of the reasons that the Past hypothesis --- this particular variant of the b-brain hypothesis ---  is widely accepted is that it is
consistent with lots of  cosmological data. In contrast, none of that cosmological data supports 
the original version of the b-brain hypothesis, or the $500$ years-ago variant
of the b-brain hypothesis. The conclusion is a simple one: No, you don't have to worry about the usual argument
that you are a b-brain, and yes, the usual argument for the second law is sound.

\centerline{***}

Great thanks to Jordan Scharnhorst and Wayne Myrvold for many conversations on this topic.


\providecommand{\href}[2]{#2}\begingroup\raggedright\endgroup

\end{document}